# Structure and Function of Window Glass and Pyrex

J. C. Phillips[1] and R. Kerner[2],

1. Dept. of Physics and Astronomy, Rutgers University, Piscataway, N. J., 08854-8019
2. LPTMC, University Pierre et Marie Curie -CNRS  UMR 7600, 75005 Paris, France

**Abstract**

Window glass is a ternary mixture, while pyrex (after window glass, the most common form of commercial glass) is a quaternary.  Building on our previous success in deriving the composition of window glass (sodium calcium silicate) without adjustable parameters, and borrowing from known reconstructed crystalline surfaces, we model pyrex as silica clusters with a specific ternary interface.  Our global model explains the thermal expansivity contours of ternary sodium borosilicates, and it is consistent with the optimized resistance of pyrex to mechanical and thermal shocks. It suggests new directions for studying the nanoscopic structure of these remarkable materials.

In principle the structure of glasses is an exponentially complex combinatorial problem. In the early days of glass science it was customary to dismiss this problem by saying that network glasses, for example, form "continuous random networks", or that metallic glasses could be modeled by "random" packing of hard spheres.  Modern glass theory goes far beyond these early models, but there are still many approaches to understanding glass structure and properties. One approach, that has the advantage of being close to commercial practice, is variational (standard for mathematical studies of non-polynomial complete (complex) problems) and focuses attention on optimizing structural properties as a function of composition. Here we adopt this approach to study pyrex  (next to window glass, this is the most common form of commercial glass), using the methods of surface science to study internal interfaces.

Borosilicate glasses mix tetrahedrally disposed $Si(O_{1/2})_4$ with ring-disposed $B_2O_3 = [(BOO_{1/2})_6]/3$, raising an interesting question: is it possible to characterize the network properties of borosilicates topologically? Topology is a powerful tool for analyzing the structure of network glasses, as proved by the derivation of the ternary composition of window glass [74% $SiO_2$, 16 % $Na_2O$, 10% $CaO$] without using adjustable parameters, as shown in Fig. 1; see [1] for details. The central reason for the success of the topological approach to the structures of network glasses is that optimized glassy networks are stress-free [2] and fill space maximally according to rules obtainable from crystal chemistry [3]. These rules by themselves are incomplete (for instance, they are incapable of distinguishing quantitatively between corner-sharing and edge-sharing tetrahedra), but when combined with experimental data (Raman, NMR, etc.), theory often confirms optimal compositions within a few % (practically exact for many cases, including window glass). Thus it appears that topology is indeed a long-sought "royal road" to understanding the structures and compositions of optimized network glasses [4].

Borosilicate glasses are of growing interest, especially for nuclear waste encapsulation [5,6], but they are already familiar both within and without laboratories, as pyrex is an optimized borosilicate glass, selected in 1915 from thousands of compositions for its optimized ability to sustain mechanical shock [7]. The ability to resist thermal shock depends on thermal expansivity, and because of thermal fluctuations during quenching there is a good correlation between the two abilities to resist shock [8]. However, we will also find that this correlation has a firm nanoscopic basis, which is important for commercial optimization, and may well be especially significant for small carefully quenched samples.

When glasses are quenched from the melt, they usually have larger volumes than the weighted molar volumes of their crystalline constituents. This "free volume" is normally distributed in the form of thin internal surface layers of lower density separating nanoclusters. In the case of ternary window glass [74% $SiO_2$, 16 % $Na_2O$, 10% CaO] [1], it was possible to ignore the internal surfaces, and to treat the network (see Fig. 1) as a stress-free mixture M with an average number of rings passing through each cation of R = 6 (just as in pure silica). The (invalid for most silicates) assumption {"prediction") [1] that the distribution of cations in soda lime silicate glasses is nearly homogeneous ("random") near the optimized composition of window glass has since been supported by NMR experiments [9], which also showed evidence for local Na-Ca pairing, in agreement with Fig. 2 of [1]. However, the shock resistance of quaternary pyrex (80.6% $SiO_2$, $SiO_2$, 13.0%$B_2O_3$, 4.1%$Na_2O$, 2.3%$Al_2O_3$) empirically requires that these internal surfaces be chemically strengthened by the elimination of CaO (cation valence Z = 2), as well as replacement of ¾ of the $Na_2O$ (cation valence Z = 1) by $B_2O_3$ (cation valence Z = 3), the addition of some silica (cation valence Z = 4), and even the replacement of 3% of the silica by $Al_2O_3$ (cation valence Z = 3, very refractory, higher melting point (2000C) than silica (1700C)). This is all very well, but how can these larger Z elements be added to the network without producing a large internal pressure, whose fluctuations would amplify external stresses, thus producing undesirable mechanical or thermal failure?

This problem is typical in the optimization of materials – when one has a material (such as ternary window glass) whose properties already appear to be optimized, it seems that any attempt at further optimization is self-defeating. In a series of elegant experiments Abe established [10] that the structure of ternary borosilicate glasses (1-x-y)$SiO2, xB_2O_3, y Na_2O$ near the pyrex composition is qualitatively different from that of all silicate glasses similar to window glass. His most dramatic result, obtained from time-temperature studies of weighted fiber elongation, was that the borosilicate glasses continued to relax locally even at viscosities as high as $10^{12}$ p, where the network structure of silicate glasses had already stabilized. He suggested that the borosilicate relaxation was related to the formation of the $B_5O_8$ clusters shown in Fig. 2. He argued further that the function of y$Na_2O$ was to catalyze the formation of these complex clusters. The function of $Al_2O_3$ was not identified, but he noted that addition of a small amount of $Al_2O_3$ enhanced the distinctive properties of borosilicates, in accordance with commercial practice.

Our proposed solution for pyrex is sketched in Fig. 3. Silica nanoclusters are formed in the melt, and their interfaces are decorated with $Na_2O$ and $B_2O_3$. (This model is a very sophisticated refinement of the "spherical coated clusters consisting of a core of one species coated by the other" described for binary hard sphere mixtures using MDS [11].) To understand the interfacial stress, it is helpful to compare pyrex to the simple prototypical binary $(Na_2O)_x(SiO_2)_{1-x}$, whose phase diagram has been very well studied [1]. The reduction in network constraints (softening, reduction of $T_g$) by the addition of $Na_2O$ (average number of resonating bonds $<R> = 4/3$) to silica ($<R> = 8/3$) is compensated by addition of oxygen bond-bending constraints, so that $(Na_2O)_x(SiO_2)_{1-x}$ forms a stress-free network between $x = 0$ and $x = 0.20$ [1,12], with the glass transition temperature $T_g$ decreasing rapidly with x.. In this simple binary mutually repulsive $Na_2O$ coats the silica cluster interfaces (right half of Fig.1).

In pyrex on the opposing face of the neighboring cluster, we replace $Na_2O$ and several (~2) layers of $SiO_2$ by $B_2O_3$. Pyrex contains ~ 3 times as many $B_2O_3$ molecules as $Na_2O$. The "extra" $B_2O_3$ is probably used to build inner "healing" layers of $B_2O_3$; these could consist entirely of tetrahedral $B(O_{1/2})_4$, as NMR has shown that these occur in borosilicates [12] (tetrahedra pack better than rings, and here the $B(O_{1/2})_4$ tetrahedra pack well with silica tetrahedra), or there could also be some $B(O_{1/2})_3$ triangles, as supposed by Abe [10]. In any case, the "healing" layers terminate with a surface monolayer of $O_{1/2}$-B=O (non-bridging O). If we fuse B=O into a single cation, then this surface layer is the charge-conjugate topological isomorph of the $O_{1/2}$-Na surface layer.

In Fig. 3 two (not just one) things favorable for shock resistance happen: first, the outer left oxygen anions are attracted electrostatically by the outer right Na cations, and second, interfacial stress is minimized, because the molar volumes of $B_2O_3$ and $Na_2O$ [which fix the respective length scales] are virtually identical (as are their molecular masses, which means that there is also little temperature dependence to the lateral misfit strain energy). In reduced units the crystalline molar volumes are: $SiO_2$ (cristobalite) = 1.00, $Na_2O$ = 1.05, $B_2O_3$ = 1.09, and $Al_2O_3$ = 0.99. Of course, one can say that the approximate matching of the molar volumes of $B_2O_3$ and $Na_2O$ is "accidental", but when a certain quaternary composition is selected from thousands of alternatives, it is just such "accidents" that are needed. Alternatively, what we have here is a kind of evolutionary selection by "trial and error", and such empirically optimized selection is hardly ever accidental.

We now discuss details of the model. For simplicity both silicon subnetwork surfaces in Fig. 3 are drawn in (100) crystalline orientation, but in the glass one expects a distribution of orientations, which should be averaged over. After such an average over orientations, scalar factors (specifically, molar volume differences) should dominate internal network interfacial (both hydrostatic and shear) misfit energies. Moreover, the interfacial structure shown in Fig. 3 refers specifically to an optimized composition near 80% silica. Thus the dominance of (1B,3Si) over (2B,2Si) in tetrahedral $^{[4]}B$ in a 55% silica alkali borosilicate environment [13], together with a larger concentration of $^{[3]}B$ in rings, merely reflects ring-tetrahedra nanoscale phase separation, with (1B,3Si) $^{[4]}B$ interfaces with (111) oriented silica interfaces. Note that the matched interfaces in Fig. 3 (80% silica) will be stabilized primarily by longitudinal (not shear) forces, which is another way of saying that the selected configuration of pyrex can be exceptionally stable both mechanically and thermally. Conversely, opposing ionic interfaces covered only by $Na_2O$ (as in $(Na_2O)_x(SiO_2)_{1-x}$) will experience isotropic Coulomb repulsive forces, with substantial shear components, and larger thermal and mechanical instabilities.

Just as the d = 3 silica clusters are separated by d = 2 internal interfaces, so stress accumulation (or alternatively, quenching kinetics) separates the d = 2 internal interfaces into patches separated by d = 1 line segments. Refractory alumina edge-sharing tetrahedral chain segments presumably reinforce these edges, thereby further enhancing pyrex's ability to sustain mechanical or thermal shock. Quite generally one can argue that in an isotropic medium (surface/volume) ~ (line/surface); this relation can be used to place limits on the $Al_2O_3$ concentration. An upper limit $f_2$ is obtained if "surface" is interpreted to mean all components except silica, so that (surface/volume) ~ 0.2, giving $f_2$ ~ 4%. Alternatively, one could include only the outer $B_2O_3$ and $Na_2O$ layers, so that (surface/volume) ~ 0.1, giving a lower limit $f_1$ ~ 1%. The actual value for $Al_2O_3$ in pyrex is 2.3%, comfortably between $f_1$ and $f_2$.

We can gain further insight into the structure of pyrex by analyzing the results of a mean-field model with modest correlations, similar to the window glass model [1]. We can look upon the pyrex network as being composed of four elementary building blocks: Si $(O_{1/2})_4$ [**A**], B$(O_{1/2})_3$ [**B**] Na $(O_{1/2})$ [**C**], and Al $(O_{1/2})_3$ [**D**]. Only the A-block represents an equivalent of one $SiO_2$ molecule, while the other three (B,C and D blocks) represent only halves of corresponding molecules. This is why the concentrations of these blocks, denoted by $\mathbf{p}_A$, $\mathbf{p}_B$, $\mathbf{p}_C$ and $\mathbf{p}_D$ are not the same as the chemical composition denoted by (1-x-y-z), x, y and z, respectively. If the total number of molecules is N, then the numbers of various blocks will be: N(1-x-y-z) (for the A blocks), 2Nx (for the B blocks), 2Ny (for the C-blocks), and 2Nz (for the D-blocks), because each molecule (except for $SiO_2$) gives rise to **two** building blocks. Therefore the concentrations of blocks A,B,C and D can be expressed as functions of chemical concentrations (1-x-y-z), x, y and z as follows: $\mathbf{p}_A$ = (1-x-y-z)/(1+x+y+z), $\mathbf{p}_B$ = 2x/(1+x+y+z), $\mathbf{p}_C$ = 2y/(1+x+y+z), and $\mathbf{p}_D$= 2z/(1+x+y+z). For the pyrex composition (80.6% $SiO_2$, 13.0% $B_2O_3$, 4.1% $Na_2O$ and 2.3% $Al_2O_3$) the corresponding block concentrations will be: $\mathbf{p}_A$ = 67,5%, $\mathbf{p}_B$ = 21,8% , $\mathbf{p}_C$ = 6.9% , and $\mathbf{p}_D$ = 3.8%.

In order to fix the optimal values of concentrations (x,y,z) one needs three independent equations. One of the three equations can be readily derived from the maximal topological homogeneity applied to the ring structure, i.e. the medium-range order, as in window glass [1]. There the reduction in average number of rings due to Na creating non-bridging oxygens is compensated by refractory Ca "zipping together" the silica rings attached to it, increasing the number of rings by 15. Here a similar count for Al-centered tripods "zipping together" surrounding A-blocks creates 30 excess rings created around each D-block. Supposing that each B-block behaves in fact in the same way as a C-block i.e. "eats out" one of the oxygen bonds stemming from an A-block (Si-centered tetrapode), thus reducing the number of rings by three, one should have 10 times as many B and C blocks taken together as the D-blocks: $(p_B + p_C) = 10\, p_D$. This equation is roughly satisfied: we have $(p_B + p_C) = 28.7\%$ and $p_D = 3.8\%$. [This is one of the places where the interfacial model of Fig.1 is helpful: it shows that the D $Al_2O_3$ blocks play a different structural role than the B ($B_2O_{3)}$ and C ($Na_2O$) blocks, so the equation should not be satisfied exactly. Note, however, that this equation also explains why the D $Al_2O_3$ blocks have such a large effect on the network, even though their concentration is small.]

Another equation can be produced if we take into account topological equilibrium between tetrapods and tripods. The B and C blocks transform the A-blocks (the Si-centered tetrapods often called $Q_4$ in publications discussing silicate glasses) into tripods (denoted by $Q_3$). Now, it is reasonable to assume that the $Q_4$ and $Q_3$ entities should not concentrate too much – they should "dilute" as homogeneously as possible, avoiding local pairing of one or another type. Supposing that each B or C block transform exactly one $Q_4$ into a $Q_3$, we conclude that out of (1-x-y-z) A-blocks there will be (2x + 2y) transformed into $Q_3$ blocks (with one neutralized oxygen bond out of four). If we assume that as a result, each $Q_3$ block is surrounded by three $Q_4$ blocks, avoiding direct contact between two $Q_3$ 's and that each $Q_4$ block is surrounded by four $Q_3$ 's, we arrive at the stochiometric ratio ¾. This leads to the following equation:

$$4\,( 1 - 2\, p_B - 2\, p_C - 3\, p_D ) = 3\, (p_B + p_C)$$

The third equation can be derived from the rigidity theory. Supposing that the action of $Na_2O$ eases the constraints by suppressing 3 constraints per one Na+O Group (one broken bond and two broken angular constraints), whereas the action of $B_2O_3$ via the formation of doubly bonded oxygens creates an extra constraint, we arrive at the relation

$$p_B = 3\ p_C$$

Together with the remaining two equations expressed by means of the variables $p_A$, $p_B$, $p_C$ and $p_D$ we get the following system:

$$p_A + p_B = 10\ p_C$$
$$p_B = 3\ p_C$$
$$4\ (1 - 2\ p_B - 2\ p_C - 4\ p_D) = 3\ (p_B + p_C)$$

whose solution is: $p_A = 0.650$ [0.675], $p_B = 0.239$ [0.218], $p_C = 0.079$ [0.069] and $p_D = 0.032$ [0.038], which agrees well with the real values [in brackets]. However, the special property of pyrex- excellent thermal and mechanical stability, unlike $(Na_2O)_x(SiO_2)_{1-x}$, is favored by the special interfacial attraction illustrated in Fig. 3.

While the experiments of Abe were classic more than 50 years ago, today we have at our disposal many probes of microscopic structure that could be used to distinguish between the models shown in Figs. 2 and 3. One of the most sophisticated probes is magic-angle spinning NMR. The doubly bonded interfacial B is unlikely to produce an observable NMR signal, but it is quite possible that the tetrahedral $^{[4]}$B populations [13,14] could exhibit a large non-linearity in composition (possibly a bulge, or even a peak) near the pyrex composition. Further progress in understanding pyrex may be obtained by extending earlier work on $^{[4]}$B in a 55-65% silica alkali borosilicate environment [13] to the pyrex neighborhood of 80% silica, and testing our idea that the attraction between B=O and Na maximizes viscosity of the melt just above $T_g$ (Fig. 4 of [10]) and explains high shock resistance.

However, the large qualitative differences between the window glass and pyrex families suggest a more direct method, namely precise study of boson peaks in these two families, which have been studied intensively primarily for silica [15] and boroxyl [16] glasses. The clusters responsible for the boson peak in silica typically contain ~ 1000 atoms [15], and hyper-raman scattering suggests that the boroxyl clusters are also large [16]. Our models predict that the departure from linearity [17] of the composition dependence of the boson peak in window-glass like silicates will be unusually weak, while that in pyrex-like borosilicates could still be relatively strong (several %, possibly 10%). Finally, the (1-x-y)SiO2,xB$_2$O$_3$,y Na$_2$O ternary phase diagram for thermal expansivity [Fig. 6 of 10] shows a trough between C (x = 0.40, y = 0) and D (x = 0.85, y = 0.15). We believe that this trough is associated with a frustrated crossover from a 3-d morphology (small expansivity) to a layered 2-d morphology (large expansivity due to weak interlayer forces). The properties of this trough are best studied with alloys that cross it symmetrically at (C + D)/2 [x + y = 0.70], rather than the often-used ones that start from C [x + y = 0.40], and observe only one side of the trough.

In conclusion, we have proposed a new structural model for pyrex (next to window glass, the most common form of commercial glass), suggested new experimental directions, and made specific, easily tested predictions. The unique properties of pyrex, which have led to its enormous practical successes, are modeled as the result of optimization of internal interfaces and edges between silica-like nanoclusters [15]. Many alternative structural models of glasses are non-variational in nature, and rely on nucleation concepts that idealize the glass transition as nearly first-order. These models involve mean-field concepts (such as surface tension) that are not easily quantified chemically. Our model, which focuses on the variational properties of stress-free, or stress-minimized networks, readily leads to specific chemical models. In the case of window glass, the mean field equations are essentially exact, while they predict the composition of pyrex with errors of ~ 2%. It is just these very small differences, largely inaccessible to most theories, that require a designed cluster model and are responsible for the distinctive physical properties of pyrex. Moreover, network models are of great interest as paradigms of possible approaches to exponentially complex problems in general, in fields as far removed as high temperature superconductivity [18] and solvability of large-scale Boolean algebras in computer science [19-21].

We are very grateful to P. Richet for showing us [10], and to K. Rabe for a stimulating conversation.

# Figure Captions

Fig. 1. Topological derivation of the composition of window glass. The mean-field (stress-free) condition is represented by the line M, while the ring condition is represented by the line R. The two lines intersect at W, the composition of window glass. See [1] for details.

Fig. 2. Two-dimensional projection of $B_5O_8$ cluster based on a central tetrahedron decorated by four triangles, from Fig. 7 of [10].

Fig. 3. Suggested interfacial structure for pyrex-like $SiO_2 – B_2O_3 – Na_2O$ interface. The network structure near the interface exhibits several favorable features that would optimize the mechanical shock resistance of the glass and maximize the viscosity of the deeply supercooled liquid.

Fig. 1.

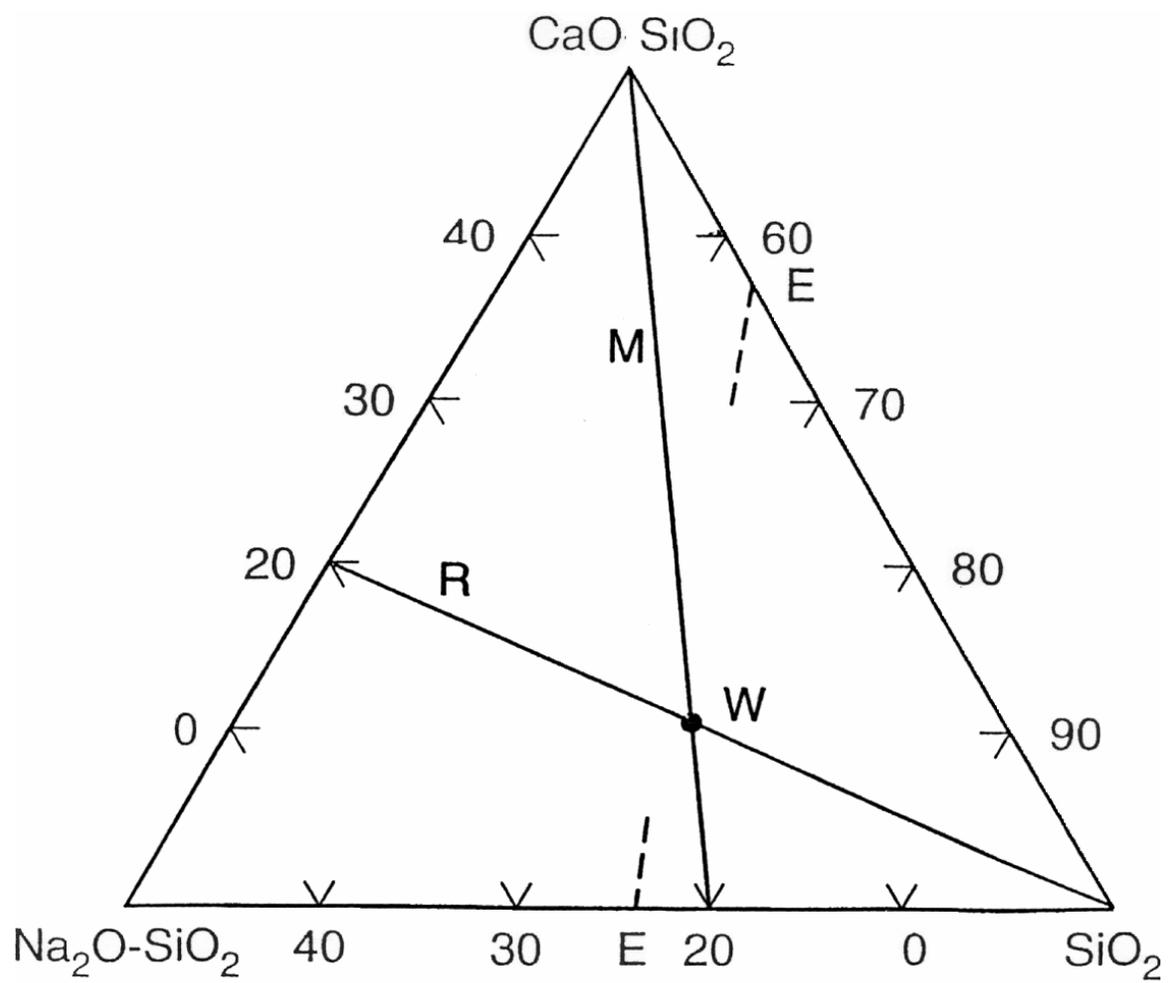

Fig. 2.

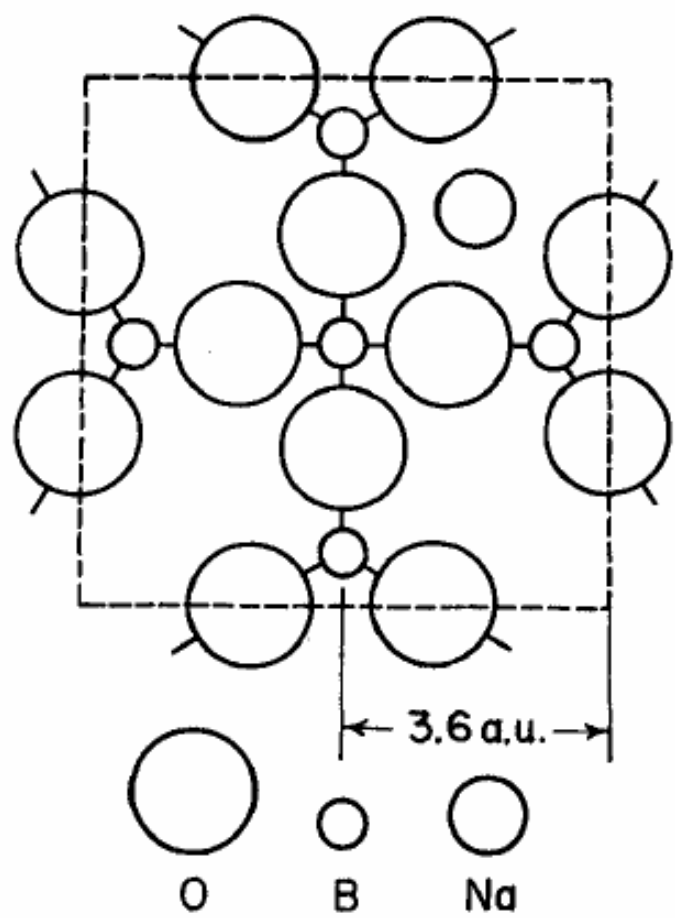

Fig. 3.

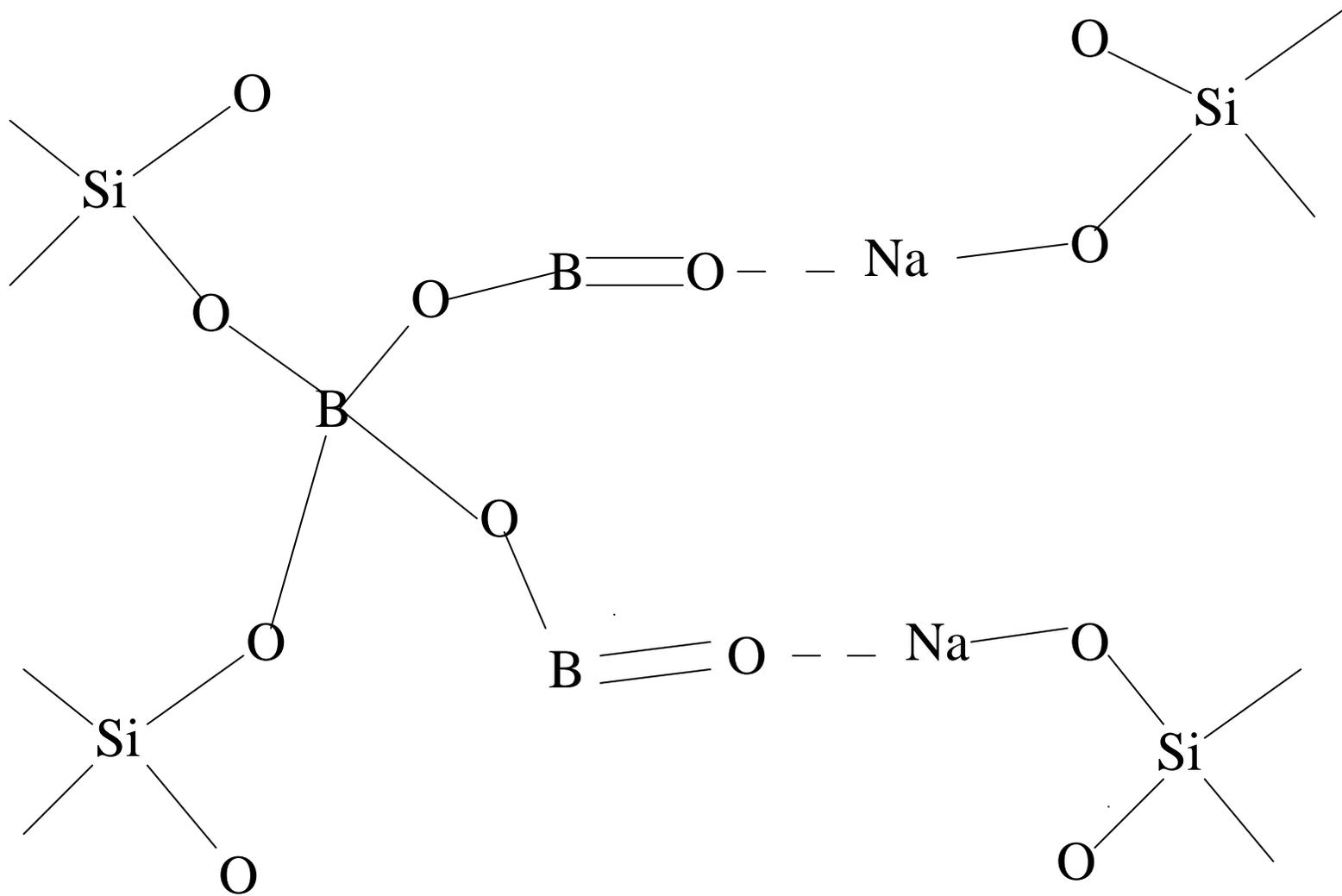